\documentclass[%
 reprint,
 superscriptaddress,
 amsmath,amssymb,
 aps,
 pre,
 floatfix,
]{revtex4-1}

\usepackage{graphicx} 
\usepackage{footnote}
\usepackage{footmisc}
\usepackage{xcolor}   

\usepackage{lipsum} 
\usepackage{dcolumn}  
\usepackage{bm}       
\usepackage{hyperref} 
\hypersetup{          
 pdftitle={},
 pdfauthor={Ohad Shpielberg},
 pdfsubject={},
 pdfkeywords={},
 unicode=true,
 breaklinks=true,
 linkcolor=blue, 
 citecolor=[rgb]{0.0,0.7,0.1},
 pdfborder={0 0 0},
 pdfborderstyle={},
 colorlinks=true,
 bookmarksnumbered=true,
 bookmarksopen=true,
 bookmarksopenlevel=1,
}

\usepackage[all]{xy}                        
\SelectTips{cm}{10}                         
\usepackage[abbreviations,xlatin]{foreign}  
\defasforeign[aesthetic]{\ae{}sthetic}      

\usepackage[normalem]{ulem}                 
\usepackage{comment}                        
\begin{document}


\title{Universality in dynamical  phase transitions of diffusive systems}
%

\author{Ohad Shpielberg}
\email{ohad.shpilberg@college-de-france.fr}
\affiliation{Laboratoire de Physique Th\'{e}orique de l'\'{E}cole Normale Sup\'{e}rieure de Paris, CNRS, ENS  \& PSL Research University,UPMC  \& Sorbonne Universit\'{e}s, 75005 Paris, France.}

\author{Takahiro Nemoto}
\email{nemoto@lpt.ens.fr}
\affiliation{Philippe Meyer Institute for Theoretical Physics, Physics Department, \'Ecole Normale Sup\'erieure \& PSL Research University, 24 rue Lhomond, 75231 Paris Cedex 05, France}

\author{Jo\~ao Caetano}
\email{jd.caetano.s@gmail.com}
\affiliation{Laboratoire de Physique Th\'{e}orique de l'\'{E}cole Normale Sup\'{e}rieure de Paris, CNRS, ENS  \& PSL Research University,UPMC  \& Sorbonne Universit\'{e}s, 75005 Paris, France.}

\date{\today} 


\begin{abstract}
 Universality, where microscopic details become irrelevant, takes place in thermodynamic phase transitions. The universality is captured by a singular scaling function of the thermodynamic variables, where the scaling exponents are determined by symmetries and dimensionality only.   Universality can persist even for non-equilibrium phase transitions. It implies that a hydrodynamic approach can capture the singular universal scaling function, even far from equilibrium. 
In particular, we show these results for phase transitions in the large deviation function of the current in diffusive systems with particle-hole symmetry.   For such systems, we find the scaling exponents of the universal function and show they are independent of microscopic details as well as boundary conditions.

\end{abstract}

\maketitle




\section{Introduction}
 A long standing goal in the study of non-equilibrium is to generalize and implement the vast knowledge accumulated in the study of thermodynamic phase transitions \citep{Zinn-Justin_book,Yeomans,Sachdev}. In equilibrium, the relevant thermodynamic potential, e.g. the free energy, becomes non-analytic at the 
transition point. For a continuous phase transition, the
thermodynamic potential is composed of a regular part and a singular universal part -- a scaling function of the relevant thermodynamic variables. The scaling function is characterized by critical exponents, which in turn classify
the physics into universality classes that depend only on the symmetry and dimensionality of the model.

Non-equilibrium systems are generally sensitive to microscopic details, boundary conditions and initial conditions. Therefore, it is appealing to find where can universality take over in non-equilibrium systems, from both a theoretical and a practical viewpoint. If universality takes over, it is tempting to assume that a coarse grained (hydrodynamic) theory can capture the singular universal behavior. The purpose of this paper is to show that this is indeed the case for an analytically tractable setup.    

It has been suggested long ago to build a thermodynamic formalism for non-equilibrium systems by looking at probabilities over time realizations rather than looking
at the instantaneous energy states \citep{Ruelle}. To illustrate this idea, let us consider two particle reservoirs, coupled through a $1D$ transport channel of size $L$ -- a
common non-equilibrium setup. The hallmark of non-equilibrium in such systems is a non-vanishing current. For this reason, a natural quantity of interest is $P_t (Q)$,
the probability to observe a transfer of $Q$ particles in the system during the time interval $\left[0, t\right]$. For $t \gg 1$, the probability to observe an atypical particle transfer, i.e. different than the steady state, is usually exponentially
unlikely. Thus, the large deviation function (LDF)
is defined by the  function $I(J) = -\frac{1}
{t} \log P_t(Q)$ for $J = Q/t$ -- the atypical mean current.    Starting from the discovery of fluctuation theorems  in the 90'  \cite{Jarzynski1997,Gallavotti1995prl}, LDFs have played an important role in the modern development of non-equilibrium
theories \cite{Touchette2009}. Since the LDF constrains the system to exhibit a mean atypical current $J$, we can define
an associated  mean spatio-temporal particle occupancy in the system, where the mean is over all spatio-temporal evolutions that support the particle transfer $Q$~\cite{Chertrite2013}.

Similarly to thermodynamic phase transitions, dynamical phase transitions (DPTs) are defined as non-analytic points in the LDF. A variety of DPTs are identified in a broad range of non-equilibrium systems, such as in high-dimensional chaotic chains \cite{tailleur2007probing,bouchet2014,laffargue2013}, kinetically constrained glass models \cite{garrahan2007,Hedges2009,Pitard2011,Speck2012,Bodineau2012a,Limmer2014,Nemoto2017prl} or active self-propelled particles \cite{Cagnetta2017,Whitelam2018,nemoto2018optimizing}. The transition is manifest in e.g. a dramatic change in the mean spatio-temporal particle occupancy \citep{Bertini2005,Appert-Rolland2008,Bodineau2005,Bodineau2007,Baek2016b,Shpielberg2017b}. 
In this paper, we especially consider 1D diffusive processes that are symmetric to the exchange of particles and vacancies.  In this case, it is known that the observed particle occupancy becomes independent of both space and time  in a range determined by the critical value $J_C$ \citep{Appert-Rolland2008,Bodineau2005,Bodineau2007,Baek2016b,Shpielberg2017b}. We show that the singular part of the LDF is universal, irrespective of microscopic details and boundary conditions.  Namely $I(J)= I_{\textnormal{reg}}+I_{\textnormal{sing}}$, where 
\begin{equation}
I_{\textnormal{sing}} = \frac{1}{L^{2+\alpha}}{\phi}(\delta u L^\beta) \label{eq:singular exponents}
\end{equation}
such that $\delta u $ is a universal parameter that vanishes as $J \rightarrow J_c$.

\begin{figure*}
\begin{center}
\includegraphics[width=0.32\textwidth]{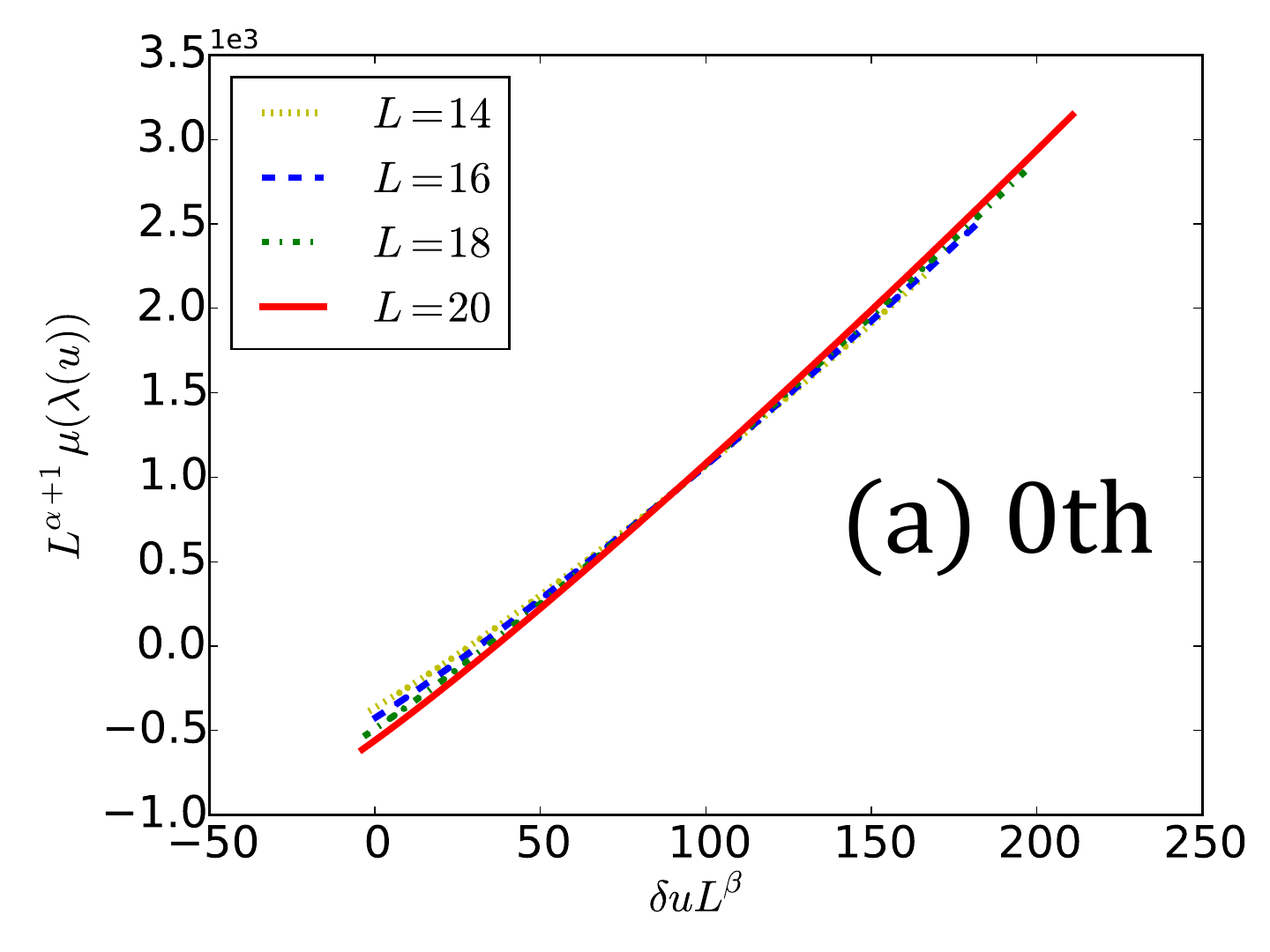} 
\includegraphics[width=0.32\textwidth]{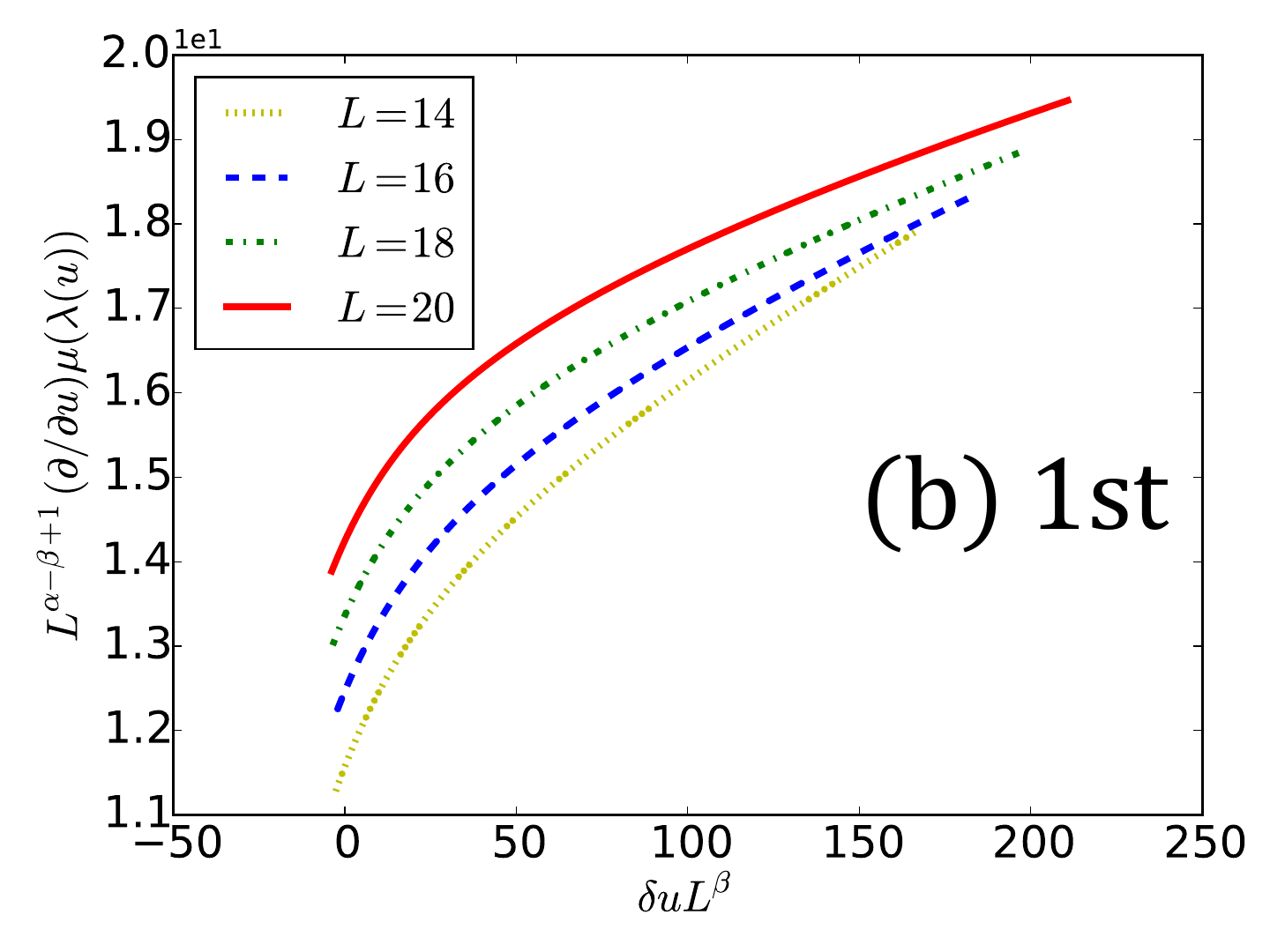} 
\includegraphics[width=0.32\textwidth]{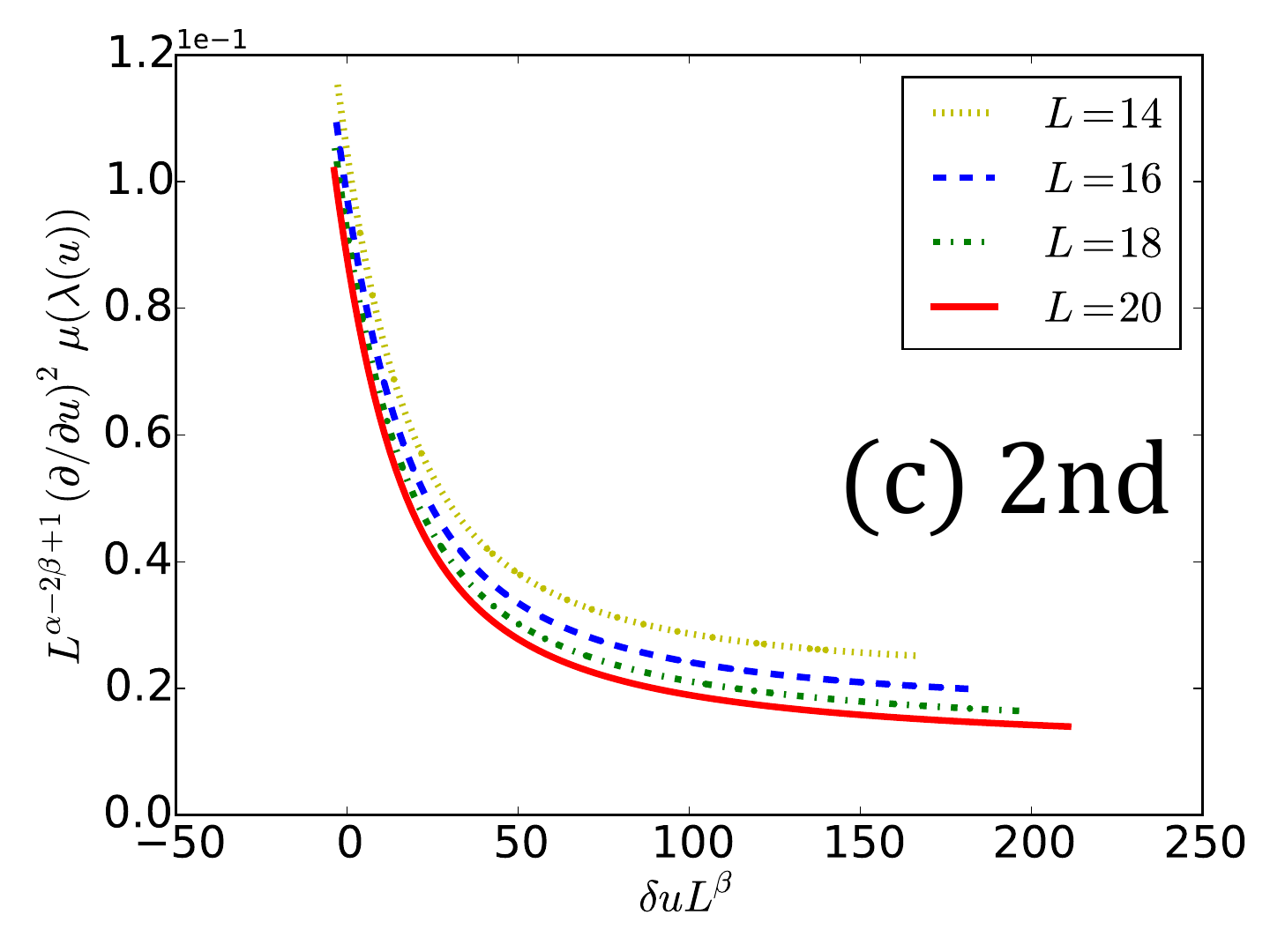} 
\includegraphics[width=0.32\textwidth]{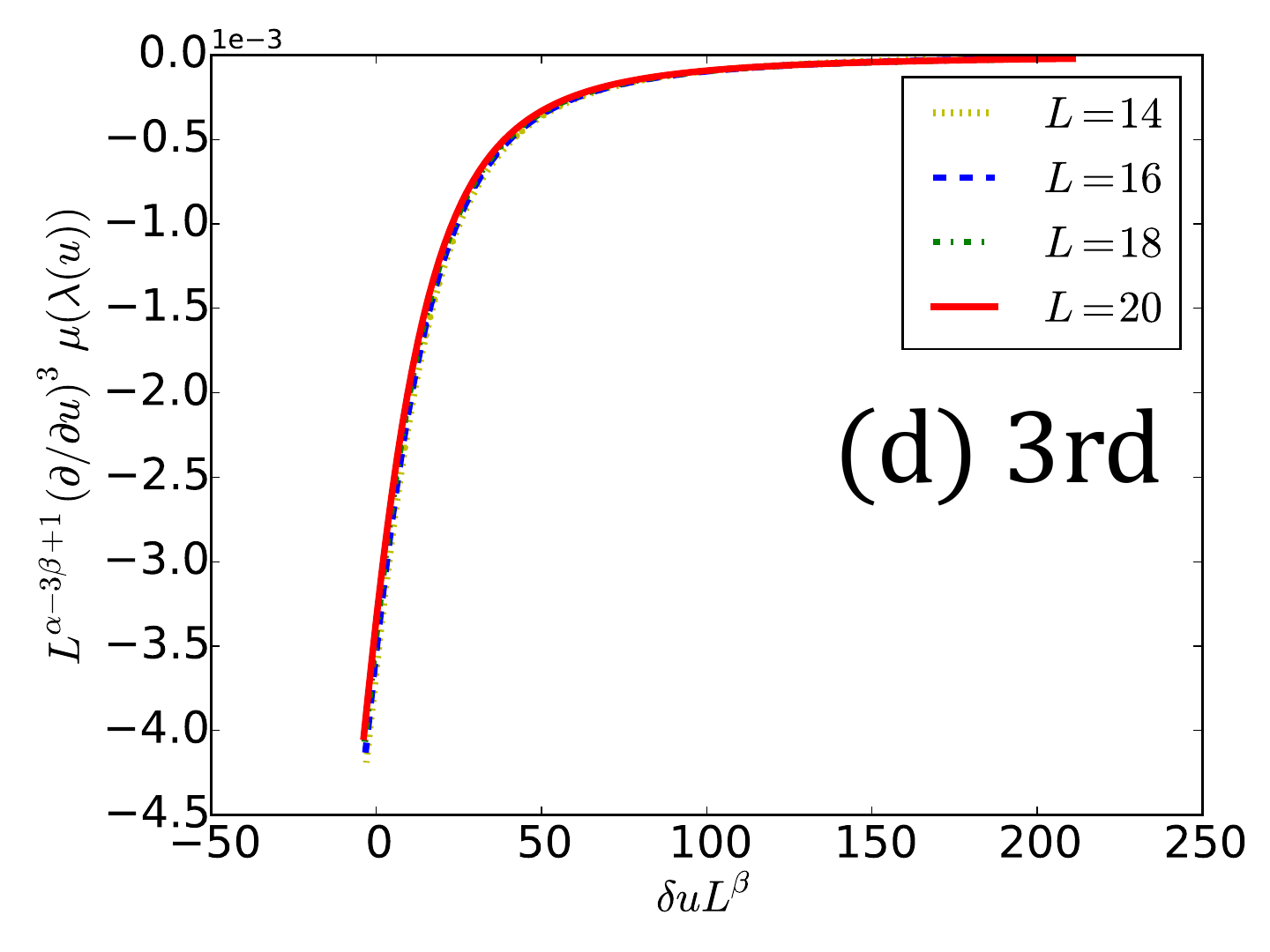} 
\includegraphics[width=0.32\textwidth]{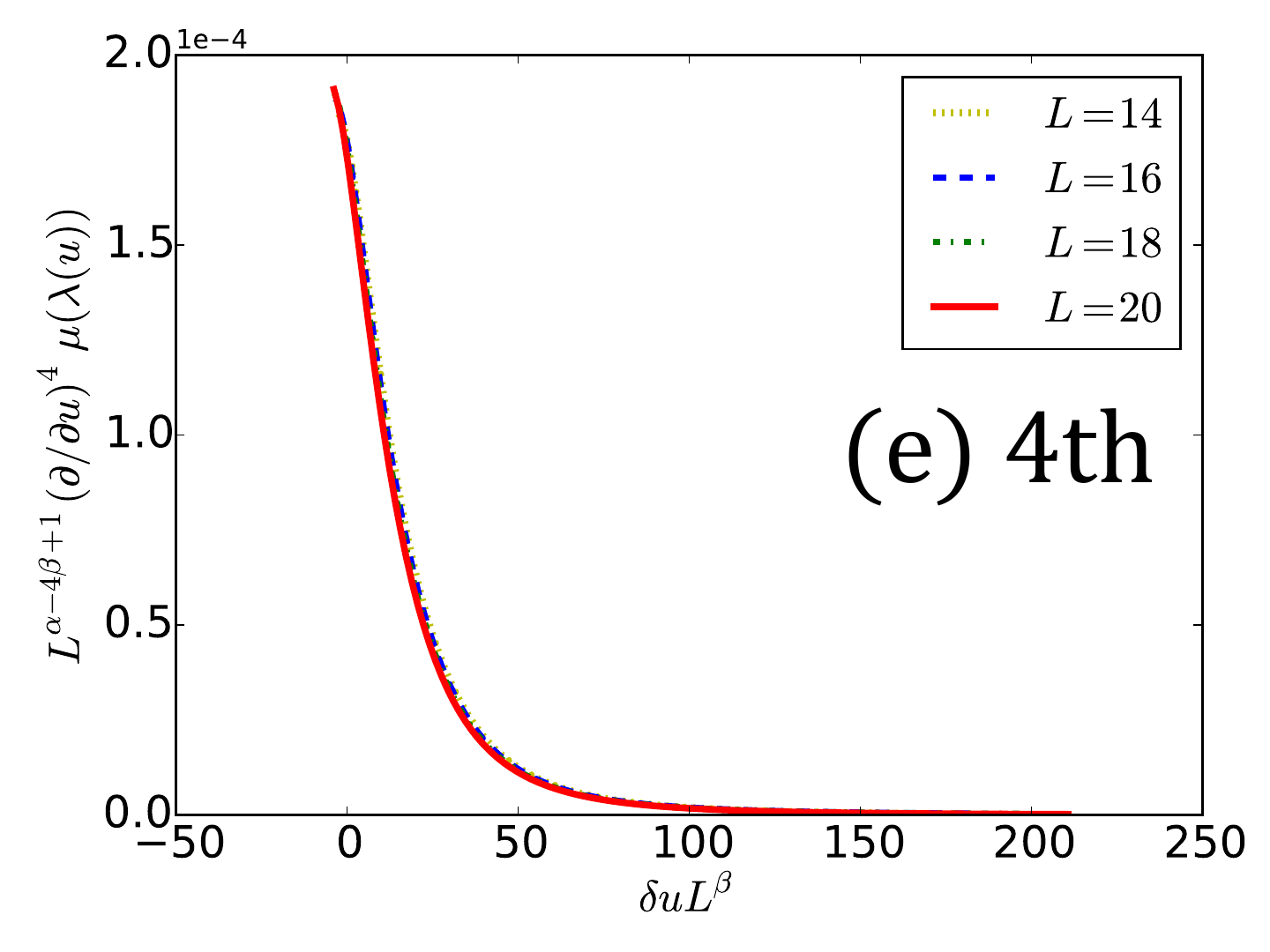} 
\includegraphics[width=0.32\textwidth]{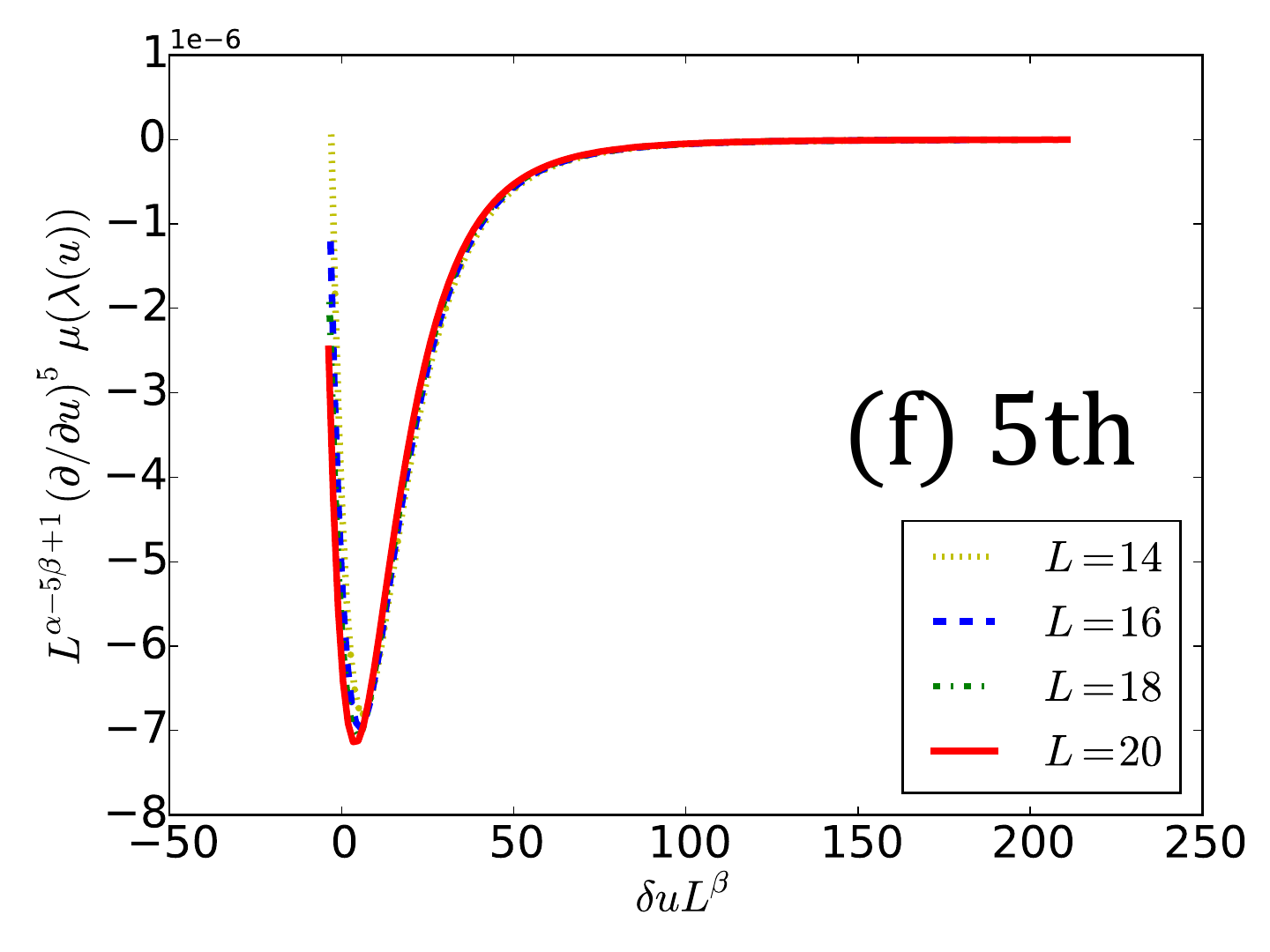} 
\caption{\label{fig:third_order} 
The $d$-th derivative with respect to the universal parameter $u$ of the cumulant generating function  $\mu(\lambda)$ and a proper rescaling
with $L^{\alpha-d\beta }$ as a function of $\delta u L^\beta$
is plotted for the weakly asymmetric exclusion process (WASEP) on a ring \citep{Mallick2015,Chou2011,Derrida2007}. The cumulant generating function $\mu(\lambda)$, as defined in the text, is a Legendre-Fenchel transform of the LDF and carries the same scaling behavior for the singular term.  For the third derivative, the universal function becomes dominant over the non-universal part even for a relatively small system size $L$. The convergence to a scaling function $\tilde \phi(r)$ is convincing already for small systems $L=14,16,18,20$ with $\alpha = 1/3,\beta = 2/3$ for the $d=3$ derivative. 
As mentioned in Section~\ref{sec:numerical verification}, the singular point $r=0$ may be shifted, especially for small systems. 
In the WASEP, as detailed in Appendix~\ref{sec:app:smatrix}, each site is occupied by at most one particle. Particles hop to a left or right empty neighbor with a $\exp(1\pm E/L)$ rate, where we take here $E=10$. To get this figure, we diagonalize the corresponding biased matrix~\cite{Lebowitz1999,Garrahan2009} numerically. See \ref{sec:numerical verification} for more details. }
\end{center}
\end{figure*}

In order to exhibit the universality and find the scaling exponents $\alpha,\beta$, we employ analytical tools as well as corroborating results using numerical analysis. First, we use the macroscopic fluctuation theory (MFT). The MFT is a hydrodynamic theory of diffusive systems. It was used to obtain various results, e.g. current fluctuations, non-equilibrium fluctuation induced forces, escape times of interacting particles, statistics of tagged particles in single-file diffusion  \cite{Agranov2018,Aminov15,Krapivsky2015,Bouchet2016,Akkermans2013,Bodineau2004,Bodineau2008,Tizon16a} and many more 
\cite{Hurtado2011b,Prados2011,Prados2012,Prados2013,Lasanata2016}. The predictions are exact, up to $1/L$ corrections. The second approach relies on an exact solution of a microscopic model -- the simple symmetric exclusion process (SSEP) \citep{Mallick2015,Chou2011,Derrida2007} via the Bethe ansatz. The Bethe ansatz 
allows to determine the energy eigenstates of many-body integrable quantum systems \cite{korepin_bogoliubov_izergin_1993}
as well as evaluating current statistics of non-equilibrium systems \citep{izergin1984,Derrida2004,DeGier2005}. The SSEP is an important model in the study of classical and quantum non-equilibrium systems \cite{Bernard2018,Kamenev2011,Derrida2004,Akkermans2013}. 
By using both methods, we evaluate $1/L^2$ corrections. Close to the transition, the leading singular behavior allows to obtain  the scaling exponents $\alpha,\beta$. 
The singular term  in the LDF is sub-leading. However, it becomes more dominant for higher and higher derivatives of the LDF with respect to $J_C-J$ (equivalently $\delta u$). Then, as shown in Fig.~\ref{fig:third_order}, the third derivative is sufficient to capture the universal behavior.

\section{The macroscopic fluctuation theory}
To unveil the universal structure of  DPTs in non-equilibrium diffusive systems, we introduce the MFT.  
Taking the limit $t,L \rightarrow   \infty$
with the fixed diffusive scaling $t/L^2$, we  define
rescaled coordinates:  $\tau = t/L^2$ and $x\in \left[0,1 \right] $. At these
scales, the coarse grained density $\rho(x,\tau  )$ is assumed to be a smoothly varying function. Here, we focus only on
processes that conserve particles at the bulk. The
 current density $j(x,\tau  )$ allows to write the continuity
equation
\begin{equation}
\partial_\tau \rho = -\partial_x j.
\label{eq:continuity equation}
\end{equation}
At the steady state, diffusive processes satisfy Fick's law
$j = \mathfrak{J}(\rho)$, where $\mathfrak{J}(\rho) = -D(\rho)\partial_x \rho + \sigma(\rho) E$. 
For a vanishing field $E$, Fick's law in \eqref{eq:continuity equation} gives the steady state diffusion equation with $D$ the diffusion.  The conductivity $\sigma$ is a measure of the response to an external field E. Generally, $D,\sigma$  are density-dependent.

The fluctuating hydrodynamics approach posits that
Fick's law can be extended to a dynamical Langevin equation \begin{equation}
j(x,\tau) = \mathfrak{J}(\rho(x,\tau)) + \sqrt{\frac{\sigma(\rho(x,\tau))}{L}} \xi(x,\tau),
\label{eq:Langevin eq}
\end{equation}
where $\xi(x,\tau)$ is a Gaussian white noise.  
The strength of the noise in diffusive systems $\sqrt{\sigma/L}$ is tuned to be consistent with the Einstein relation \cite{Derrida2007}. The dynamics of diffusive systems is thus expressed through $D$ and $\sigma$ only.  

From the Langevin equation \eqref{eq:Langevin eq} and by using  the Martin-Siggia-Rose formalism \cite{Martin1973}, the fundamental result of the MFT is derived. Namely, we find that 
the probability to observe a history $\lbrace \rho,j \rbrace$ of the system during time $\left[0,T=t/L^2\right]$ is given by 
\begin{eqnarray}
\label{eq:fund eq}
\mathcal{P} (\lbrace \rho,j \rbrace) & \sim & \exp \left(-L \, \mathcal{I}_{\left[0,T\right]}(\rho,j) \right),
\\ \nonumber
\mathcal{I}_{\left[0,T\right]}(\rho,j) &=&  \intop^1 _0 dx \intop^T _0 d\tau \, \frac{(j+D\partial_x \rho-\sigma E)^2}{2\sigma},  
\end{eqnarray}
where the continuity equation is implicitly assumed. The MFT becomes exact for $L\rightarrow \infty$.
Indeed, trying to extract microscopic details from a hydrodynamic theory is usually an ill fated attempt. 
Notice that any observable obtained through \eqref{eq:fund eq} will be dominated by the  saddle-point since $L$ is large.

It is convenient to define the rescaled LDF $\Phi(J)=L I(J)$, so that $\Phi$ is $L$ independent to leading order. Then, using the MFT to calculate the  LDF of diffusive systems boils down to solving the minimization problem
\begin{equation}
\Phi(J)= \frac{1}{T} \min_{j,\rho} \mathcal{I}_{\left[0,T\right]}(\rho,j),
\label{eq:LDF formal}
\end{equation}
where  $\lbrace \rho,j \rbrace$ satisfy the continuity equation \eqref{eq:continuity equation} and the  macroscopic particle transfer $Q = L^2 \int dx d\tau j  $ for a large  diffusive time $T \gg 1$ \footnote{We implicitly assume that particles do not accumulate in the system. See \cite{Hirschberg2015} for a contrary case. }. Since we consider non-equilibrium processes, boundary conditions  usually strongly impact the results. Here, we consider  periodic boundary conditions and boundary driven processes -- where the system is coupled to two particle reservoirs with densities $\rho_{l,r}$ respectively. The reservoirs' state is assumed to be unaffected by the interaction with the system. 
For a periodic system, the integrated number of particles is conserved. Therefore, one requires  that $\int dx \, \rho(x,\tau)$ is fixed for any $\tau$. Moreover, $\rho(x=0,\tau)=\rho(x=1,\tau)$. For boundary driven processes, $\rho(x=\lbrace 0,1\rbrace ,\tau  )$ is fixed to the boundary values
 $\rho_{l,r}$. 

\section{Dynamical phase transitions}
Finding a solution to \eqref{eq:LDF formal} is hard even for simple models. It requires  solving a partial differential non-linear problem with constraints. In \cite{Bodineau2004}, it was conjectured that the optimal density profile in current fluctuations is time-independent, which is the so-called  additivity principle. For $1D$ systems, it implies $j(x,\tau)=J,\rho(x,\tau)=\rho(x)$. The particle transfer and continuity constraints are relaxed and 
the variational principle (\ref{eq:fund eq}) is simplified to 
\begin{equation}
\Phi_{\textnormal{AP}}(J) = \min_{\rho(x)} \int dx \, \frac{(J+D\partial_x \rho - \sigma E)^2}{2\sigma}. 
\label{eq:AP LDF}
\end{equation}
This is clearly a significant improvement, as the solution of \eqref{eq:AP LDF} requires solving a non-linear ordinary differential equation \citep{Bodineau2004,Imparato2009,Bertini2005}. See \cite{Shpielberg2016,Bertini2005,Bertini2006} for discussions on the validity of the additivity principle. 
However, even for a time-independent density profile, the solution need not be unique which will usually give rise to a DPT. 
For periodic systems,  translational symmetry suggests a spatial invariant density profile so that $\rho(x) \rightarrow \rho $ which fixes the solution. This constant solution can be overtaken by a traveling wave solution as was shown in \cite{Bodineau2005,Appert-Rolland2008,Espigares2013}, amounting again to a DPT.


\section{Finite size corrections}
From now on, we focus only on models with dynamics  that satisfy particle-hole symmetry. This implies that odd derivatives of $D,\sigma$ w.r.t. $\rho$ vanish at  $\rho=1/2$. The solution of the LDF to \eqref{eq:AP LDF}  of periodic boundary conditions with mean  density $\rho=1/2$ as well as a boundary driven process with $\rho_{l,r} = 1/2$ clearly bears a special symmetry. Assuming the additivity principle, the constant density solution $\rho(x)=1/2$ is a solution for any $J$ which results in $\Phi_{\textnormal{AP}}(J) = J^2 / 2\sigma$. From here on out, $D, \sigma$ and their derivatives are always evaluated at $\rho=1/2$.  
Taking small fluctuations around this solution, namely $\rho(x,t) \rightarrow 1/2 + \delta \rho   $ and $j(x,t) \rightarrow J + \delta j   $ allows to explore the finite size corrections and whether the solution is indeed optimal. Note that $\delta \rho,\delta j$ have to satisfy the continuity equation and the integrated current constraint. For a boundary driven case the fluctuations can be recast using the Fourier representation   
\begin{eqnarray}
\delta \rho &=& \frac{1}{2}\sum_{k,\omega} k \sin (kx) (a_{k,\omega} e^{i\omega \tau} + a^\star _{k,\omega} e^{-i\omega \tau}), 
\\ \nonumber
\delta j &=& \frac{1}{2} \sum_{k,\omega} i\omega \cos (kx) (a_{k,\omega} e^{i\omega \tau} - a^\star _{k,\omega} e^{-i\omega \tau}). 
\label{eq:Foureir rep}
\end{eqnarray}
where $k=\pi n$ and $\omega = \frac{2\pi}{T}m$ for $n,m\in \mathbb{Z}$. Moreover, $a^\star _{k,\omega} = a _{k,-\omega} $ and  $a _{-k,\omega} = a _{k,\omega} $. We remark that the $k,\omega$ values have a finite cutoff of the order   $\lvert k \rvert \sim L$ and $\lvert \omega \rvert \sim L^2$ for the hydrodynamic theory to be valid. Let us keep that in mind, and set these cutoffs by $k_{\max},\omega_{\max}$ (for periodic boundary condition a slightly different representation is required, see \cite{Appert-Rolland2008}). 

To find  $\Delta \Phi \equiv \Phi - \Phi_{\textnormal{AP}} $, we rescale $a_{k,\omega}\rightarrow a_{k,\omega}/\sqrt{L}$ and obtain a perturbative Landau-like theory as
\begin{equation}
\Delta \Phi  = 
 -\frac{1}{T L} \log 
 \prod_{k\geq 0,\omega\geq 0} \int d^2 a_{k,\omega} e^{-\int dx d\tau \, \sum_{j\geq 2} \frac{S_j}{L^{1-j/2}}}
\end{equation}
with the Gaussian term $S_2 = \sum_{k,\omega} f (k,\omega)  \lvert a_{k,\omega} \rvert ^2
$ such that $f = \frac{\omega^2}{2\sigma}  +\frac{D^2}{2\sigma}  k^2 (k^2 -2u) $  with $u= \epsilon \frac{J^2 - E^2 \sigma^2  }{16D^2 \sigma}\sigma''  $ and $\epsilon = 4(1) $ for boundary driven (periodic) systems. The higher order terms $S_i$, explicitly detailed in the appendix \ref{sec:app:driven corr }, were considered here only for the boundary driven case.

Evaluating $\Delta \Phi$ boils down to performing a perturbation theory for Gaussian integrals. Let us define $\Delta \Phi = \sum_{j=1,2,...}L^{-j} \Phi_j $. We then find 
\begin{equation}
\Phi_1 =  d D \mathcal{F}(u) + c J^2 ,
\label{eq:Phi1}
\end{equation}
where $c$ is a constant that depends on the cutoffs and cannot be evaluated from a hydrodynamic theory, $d=\frac{1}{8} \, (1)$ for boundary driven (periodic) systems and $\mathcal{F}$ is 
\begin{equation}
\mathcal{F}(u) = -4 \sum_{n=1,2,...} n \pi \sqrt{n^2 \pi^2 -2u } - n^2 \pi^2 + u,  
\label{eq:F singular}
\end{equation}
already recovered in this context \citep{Appert-Rolland2008,Imparato2009,Baek2018} as well as others \cite{Beisert:2005mq,Gromov:2005gp}. 
For $u= u^\star = \pi^2/2$, $\mathcal F(u)$ is non-analytic and its derivatives diverge. This singularity has been discussed as the onset of a DPT ~\citep{Appert-Rolland2008,Imparato2009,Baek2018}. It also implies the break down of the perturbation theory close to the transition point (see appendix \ref{sec:app:non-convexity},\ref{sec:app:periodic corr}), i.e., all the higher order perturbation coefficients $\Phi_j$ diverge at this point. The singular part of the $1/L$ correction is universal -- independent of microscopic details and fully captured by the macroscopic $D,\sigma$. Therefore, it is natural to assume that a singular universal function, just like in \eqref{eq:singular exponents}, emerges from the sum of all the singular corrections. To obtain the scaling exponents $\alpha,\beta$, it is sufficient to find the dominant singular behavior of $\Phi_2$, as shown below.

The $1/L^2$ correction is cumbersome and littered 
with non-universal terms, depending on microscopic details (see appendix \ref{sec:app:driven corr }).
Focusing on the leading singular term and  defining $\delta u = u^\star - u$, we find that 
\begin{equation}
\label{eq:Phi2 sing}
\Phi_2 = \frac{15 \pi ^4 \left(D\sigma ''-2 \sigma  D''\right)}{16 D \delta u} + \mathcal{O} \left(\frac{1}{\sqrt{\delta u}}\right)
\end{equation}
as $\delta u \rightarrow 0$. We expect  periodic systems to yield a similar  leading term.   Let us now evaluate the critical exponents.


\section{The scaling function}
We have shown that the finite size corrections diverge at the critical point $u^\star$. 
For a continuous phase transition, we expect (to leading order) 
\begin{equation}
L \Delta \Phi (J) = \frac{1}{L^{\alpha}} \phi\left(\delta u L^\beta \right) + \textnormal{non universal terms}. \label{eq:scaling form}
\end{equation}
Here, $\phi(r)$ is the scaling function and the non-universal terms are of order 1. From \eqref{eq:Phi1}, \eqref{eq:F singular}, \eqref{eq:Phi2 sing}, 
we find that the leading singular term is of the form $
\phi_0 \sqrt{\delta u} +  \frac{\phi_1}{L \delta u } + \mathcal{O}(\frac{1}{L^2})  
$ where $\phi_{0,1}$ are constants. To keep the scaling \eqref{eq:scaling form}, we find that $\alpha = \beta/2 = 1-\beta$. This leads to the exponents $\alpha = 1/3, \beta = 2/3$.    


\section{Bethe ansatz for the SSEP}
\label{subsec:2nd order - Bethe}
To test whether the critical exponents are indeed universal, we corroborate our result by analyzing the finite size corrections of an integrable microscopic model -- the SSEP. The SSEP is defined by setting $E=0$ in the WASEP (See the caption of Fig.~\ref{fig:third_order} or the appendix \ref{sec:app:smatrix}). Macroscopically, it corresponds to $D=1,\sigma = 2 \rho (1-\rho)$. 
Note that, since $u$ is always negative in this case, the singularity of $\mathcal F(u)$ is not attained for real values of $J$. Yet it can still teach us about the formal structure of the universality by investigating the poles appearing in the perturbation coefficients.

For a technical reason, instead of trying to find the LDF $I(J)$, we consider equivalently the cumulant generating function (CGF) $G(s)= \sum_Q e^{-sQ} P_t (Q)$. Note that the CGF is a Legendre-Fenchel transform of the LDF. For diffusive processes, it is natural to define the rescaled CGF $\mu(\lambda)=L G(s)$ where $\lambda = sL$, similarly to the rescaled LDF structure $\Phi(J)$. 
For a Markov process, the CGF is the  ground state energy (lowest eigenvalue) of an operator $H$ associated to the Markov matrix \cite{Derrida2004} (see  appendix \ref{sec:app:smatrix}). This property makes the CGF appealing from both a numerical and theoretical perspectives as we shall see in the following. 

For the SSEP, the CGF $G(s)$ corresponds to the ground state energy of a quantum spin chain operator \cite{Derrida2004}
\begin{equation}
\begin{aligned} \label{eq:SSEPhamiltonian}
H = & \frac{L}{2}-\frac{1}{2}\sum_{i=1}^{L}\bigl[\cosh s \left( \sigma_{i}^{x}\sigma_{i+1}^{x}+\sigma_{i}^{y}\sigma_{i+1}^{y}\right)+\sigma_{i}^{z}\sigma_{i+1}^{z}\\
&-i \sinh s \left( \sigma_{i}^{x}\sigma_{i+1}^{y}+\sigma_{i}^{y}\sigma_{i+1}^{x}\right) \bigr]\,.
\end{aligned}
\end{equation}
Its eigen-system can be exactly determined via Bethe ansatz \cite{korepin_bogoliubov_izergin_1993}. In the coordinate  formulation of the Bethe ansatz, 
each particle is described by a plane-wave with their interactions embodied in a pairwise factorizable scattering matrix \cite{sutherland2004beautiful}. 
Considering $N$ particles on a ring of size $L$, such that mean density $\rho =N/L \in(0,1) $, the corresponding wave-function is parametrized by the complex parameters $\{\xi_i\}_{i=1}^{N}$ which are quantized according to the so-called Bethe equations
\begin{equation} 
\xi_{i}^{L} = \prod_{\substack{j=1 \\ j\neq i}}^{N}\left[-\frac{e^{s}-2\xi_i+e^{-s}\xi_i \xi_j}{e^{s}-2\xi_j+e^{-s}\xi_i \xi_j} \right]\,.
\label{eq:BAE}
\end{equation}
The eigenvalues are expressed in terms of the solutions of these equations through
\begin{equation}
G(s) = -2 N+e^{-s}\sum_{j=1}^{N}\xi_j+e^{s}\sum_{j=1}^{N}\frac{1}{\xi_j}\,.
\end{equation}
Note that the ground state is shifted here by $2N$ to obtain  $G(s)$. 
Based on Bethe ansatz, the CGF was already calculated to order $1/L$ \cite{Appert-Rolland2008}. Using an alternative method based on the Baxter equation \cite{Gromov:2005gp}, we  compute the $1/L^2$ corrections and numerically validate our results.

Generally, solving \eqref{eq:BAE} analytically is unfeasible for arbitrary finite $N$ and $L$ but in the thermodynamic limit where $N,L\rightarrow \infty$, they become tractable.
The key observation is that under the change of variables
$\xi_i = e^{s} (z_i +i/2)/(z_i - i/2)$, 
the Bethe equations \eqref{eq:BAE} become
\begin{equation} \label{eq:XXX}
\left(\frac{z_i+i/2}{z_i-i/2}\right)^{L} =e^{-\lambda}\,\prod_{\substack{j=1 \\ j\neq i}}^{N}\frac{z_i-z_j+i}{z_i-z_j-i}\;\;\;\; i=1,\dots,N.
\end{equation}
Eqs.\eqref{eq:XXX} are precisely the Bethe equations for the  twisted XXX$_{1/2}$ spin-chain with $\lambda$ playing the role of the twist \cite{Sklyanin:1988yz}. 
Finite size corrections to the spectrum of spin chains of this type are well studied. In particular, a powerful method based on the so-called Baxter equation was developed in \cite{Gromov:2005gp} and applied for the closely related $sl(2)$ spin chain. Based on the results of \cite{Gromov:2005gp}, we determine the finite size corrections of the SSEP to order $1/L^2$. Namely, we determine $\mu_0, \mu_1$ and $\mu_2$, where we have defined $\mu(\lambda) =\sum_{i=0}^{\infty} \mu_i L^{-i}$.  
The expressions $\mu_0$,  $\mu_1$  (see appendix \ref{sec:app:periodic corr})  agree with the previously obtained results \cite{Appert-Rolland2008}.
The full expression of $\mu_2$, which is one of our main results, has a long expression given in the appendix.
These theoretical predictions of $\mu_{0,1,2}$ are confirmed in the appendix  by comparing to a population dynamics algorithm \cite{Giardina2006xxx,tailleur2007probing,Hurtado2009,Giardina2011JSP,Nemoto_Bouchet_Jack_Lecomte,Nemoto2017,Ray2018PRL,klymko2018rare,Brewer2018}.

The interesting part of the $\mu_2$ arises at its strongest singularity. For illustration we consider the lowest mode in 
the appendix \eqref{eq:mu2_exact},
namely $k=1$, and we find the following singular  behaviour
\begin{equation}
\mu_2 \sim \frac{2 \pi ^4 \left(\theta ^2-1\right)}{\delta u \,\theta ^2}+\mathcal{O}\left( \frac{1}{\sqrt{\delta u}}\right)\,,
\label{eq:G2 corr}
\end{equation}
where we have introduced $\delta u \equiv \frac{1}{8}\theta^2 \lambda^2 + \frac{\pi^2}{2}$ and  $\theta = 2 \sqrt{\rho(1-\rho)}$. 
Under the continuation to complex values of $\lambda$, we find simple poles at the positions $\lambda = \pm \frac{2 i \pi }{\theta }$.
We then get the same type of singularity in $\delta u$ as in the hydrodynamics analysis, i.e., ($\mu_2 = \mathcal O(1/\delta u)$). Combined with the results of the MFT, we deduce that the $1/L^2$ correction diverges with $1/\delta u$.

\section{Numerical verification \label{sec:numerical verification}}
It is numerically hard to single-out  the singular universal function $\phi$ from the (unknown) non-universal terms. However, differentiation accentuates the singular term as detailed below.


In terms of CGF $\mu(\lambda)$, the scaling form eq.(\ref{eq:scaling form}) 
is written as
\begin{equation}
L \left [ \mu(\lambda) - \mu_{\rm AP}(\lambda) \right ] =  \frac{1}{L^{\alpha}} \tilde \phi (\delta u L^{\beta}) + {\rm non \  universal \ terms},
\end{equation}
where $\delta u = \pi^2/2 - u$ and
\begin{equation}
u(\lambda)=\mu_{\rm AP}(\lambda) \frac{\sigma^{\prime\prime}}{ 8 D^2},
\end{equation}
\begin{equation}
\mu_{\rm AP}(\lambda)=\lim_{L\rightarrow \infty}\mu(\lambda).
\end{equation}
To derive these expressions, we have used $J=-\mu^{\prime}_{\rm AP}(\lambda)$ \footnote{Note that a shift in the critical point is also possible due to finite size effects. We do not discuss this point further}.
This scaling form indicates that the higher order derivatives of $\mu(\lambda)$ with respect to $\delta u$ is dominated by the universal function $\tilde \phi$. More precisely, 
\begin{equation}
L^{\alpha - d \beta + 1}\mu^{(d)}(\lambda) \big |_{r=\delta u L^{\beta}} = \tilde \phi(r) + {\mathcal O}(L^{\alpha - d \beta + 1}),
\label{SM:scaling_}
\end{equation}
where $\mu^{(d)}$ is the $d$-th derivative of $\mu$ with respect to $\delta u$ and $r$ is the scaling variable given as $r=\delta u L^{\beta}$. We thus can see that sufficiently large derivatives $d$ (more precisely, $d=3$ given $\alpha = 1/3$, $\beta = 2/3$) allows us to neglect $\mathcal O(L^{\alpha - d \beta + 1})$. Using the method detailed in the next paragraph, we have numerically probed $\mu(\lambda)$ for the WASEP to search for the universal scaling function.
We present the plot, showing the left-hand side of (\ref{SM:scaling_}) in Fig.~\ref{fig:third_order}. One can clearly see that the function starts to overlap from the third order derivative, supporting the prediction of the scaling exponents.

To obtain $\mu(\lambda)$ and its derivatives,  we numerically diagonalize the s-biased operator $L^s_{\mathcal C,\mathcal C^{\prime}}$, whose explicit expression is detailed as (\ref{eq:appen_Lscc}) in Appendix~\ref{sec:app:smatrix}.
In order to obtain the  derivatives of the CGF in a stable manner, we use the following method:
We denote the eigenvalue equation of $L^s$ by
\begin{equation}
L^s \varsigma = G(s) \varsigma,
\end{equation}
where $\varsigma$ is the right eigenvector associated with the principal eigenvalue $G(s)$. To get the first order derivative, we numerically solve the following equation
\begin{equation}
(L^s)^{\prime} \varsigma + L^s \varsigma^{\prime} = G^{\prime}(s) \varsigma + G(s) \varsigma^{\prime}.
\end{equation}
together with the eigenvalue equation. 
Similarly, to get the second order derivatives, we add another equation $(L^s \varsigma)^{\prime\prime} = (G(s) \varsigma)^{\prime\prime}$ to these equations. Higher order derivatives can be also calculated in the same strategy. Thanks to this method, we do not have to rely on the difference method, which increases the error of the estimation of the higher-order derivatives.




\section{Discussion}
We have probed the LDF (CGF) of the current in diffusive systems using a hydrodynamical theory, a Bethe ansatz approach and numerical simulations. For dynamics with particle-hole symmetry, a singular scaling function with universal exponents is observed.  This implies that near the transition, macroscopic fluctuations dominate and hydrodynamic theories are sufficient to observe the critical behavior \cite{Baek2018,Gerschenfeld2011}. Thus, it is understood that non-equilibrium systems are prone to universality, where not only microscopic bulk dynamics, but also boundary condition details may be washed away.

Our observation leads us to conjecture a similar scaling exponents for current fluctuations in an infinite chain, starting from a step initial conditions \cite{Derrida2009,derrida2009a,Krapivsky2012}. Consider an infinite $1D$ chain, where at time $t=0$  the sites $i\leq 0$ have mean density $\rho_l$ and the sites $i>0$ have mean density $\rho_r$ with Bernoulli distribution (no correlations between the sites).  For the SSEP, the CGF was completely determined (see Eq.2 in \cite{derrida2009a}). One can notice that for $\rho_{l,r}=\frac{1}{2}$, the CGF becomes singular for the unphysical value $\lambda = \pm i \pi$, similarly to the value obtained for the boundary driven setup. From the similar structure, it is indeed appealing to conjecture that  the universal structure shown here is carried through also in the infinite chain setup as well.  

While the universality class here involves diffusive processes with particle-hole symmetry, it is temping to check whether the exponents are valid  even outside the range of validity currently considered, e.g. in models of ballistic or anomalous transport. The nonlinear fluctuating hydrodynamics theory \citep{Spohn2014} may allow to detect the universality class in these regimes.

Two more  remarks are in order for the scaling function. Notice that for the SSEP on a ring, with  the mean density $\rho=1/2$, the singular terms vanish in  \eqref{eq:G2 corr} as well as the subleading diverging terms (see appendix \ref{sec:app:periodic corr}). This does not imply that the singular behavior changes as the density is changed by an infinitesimal amount. We expect that a similar scaling will be recovered in the next leading order expansion. Secondly, as is verified for the SSEP on a ring, the diverging term is a simple pole in \eqref{eq:G2 corr} even without the particle-hole symmetry in the density. Therefore, the critical exponents do not change for periodic boundary conditions, irrespective of the symmetry. It would be interesting to find whether the scaling exponents remain the same even when the particle-hole symmetry is broken for boundary driven processes. To verify that, it is necessary to find continuous DPTs in boundary driven processes, which are analytically tractable. Unfortunately, such transitions are  not expected to support a constant density profile that enable the direct perturbation theory performed here \cite{Shpielberg2017a}.

\begin{acknowledgments}
We thank   Y. Baek, N. Gromov, O. Hirschberg,  V. Kazakov and Elsen Tjhung for fruitful discussions. We especially thank B. Derrida for many insightful remarks. OS acknowledge the  support of ANR-14-CE25-0003. The work of JC was supported by the
People Programme (Marie Curie Actions) of the European Union's Seventh
Framework Programme FP7/2007-2013/ under REA Grant Agreement No 317089
(GATIS), by the European Research Council (Programme ``Ideas''
ERC-2012-AdG 320769 AdS-CFT-solvable), from the ANR grant StrongInt
(BLANC- SIMI- 4-2011). This work was granted access to the HPC resources of MesoPSL financed by the Region Ile de France and the project Equip@Meso (reference ANR-10-EQPX-29-01) of the program Investissements d'Avenir supervised by the Agence
Nationale pour la Recherche.
This work was also granted access to the HPC resources of CINES/TGCC under the allocation 2018-A0042A10457 made by GENCI.
\end{acknowledgments}


\bibliographystyle{apsrev4-1}   
\bibliography{refs1} 


 \clearpage

\newpage
%


\appendix





\appendix

\section{The WASEP and the s-biased matrix \label{sec:app:smatrix}}

For completeness, we detail here the definition of the WASEP and the SSEP. We also discuss the s-biased ensemble that allows to recast finding the CGF as a ground state of an operator. 
Denoting the configuration of the particles by $\mathcal C = (n_i)_{i=1}^{L}$, where $n_i=1(0)$ means the site $i$ is occupied (empty), the transition rates of the WASEP  $w(\mathcal C \rightarrow \mathcal C^{\prime})$ are given by
\begin{equation}
w(\mathcal C \rightarrow \mathcal C^{\prime}) = \sum_{i=1}^{L}\left [ n_i (1-n_{i+1})e^{E/L} +  (1-n_i) n_{i+1}e^{-E/L} \right ],
\end{equation}
where we use the periodic boundary conditions $n_0 = n_L$ and $n_1 = n_{L+1}$. Note that when $E=0$, the model is reduced to the SSEP.

The CGF of the current $G(s)$ in this model (see the main text for the definition of $G(s)$) is  the largest eigenvalue of the following s-biased matrix 
~\cite{Garrahan2009}:
\begin{equation}
L^{s}_{\mathcal C^{\prime},\mathcal C} = 
w(\mathcal C \rightarrow \mathcal C^{\prime})e^{-s \hat j(\mathcal C \rightarrow \mathcal C^{\prime})} - \sum_{\mathcal C^{\prime\prime}}w(\mathcal C \rightarrow \mathcal C^{\prime\prime}),
\label{eq:appen_Lscc}
\end{equation}
where $\hat j(\mathcal C \rightarrow \mathcal C^{\prime})$ is a microscopic current, which takes 1 (0) when a particle moves to the right (left) direction in the transition $\mathcal C \rightarrow \mathcal C^{\prime}$. 

To bridge to the macroscopic description, as discussed in the main text, we define $\lambda = sL$ and $\mu(\lambda) = LG(\lambda/L)$. Then, the Legendre-Fenchel transform of $\mu(\lambda)$ corresponds to $\Phi(J)$ given as 
eq.(\ref{eq:LDF formal}) 
in the main text. Note also that $D=1$, $\sigma = 2 \rho (1-\rho)$ in WASEP. When the additivity principle  is satisfied, $\lim_{L\rightarrow\infty}\mu(\lambda)$ becomes simply a quadratic function
\begin{equation}
\lim_{L\rightarrow \infty}\mu(\lambda) = \mu_{\rm AP}(\lambda) \equiv -E \sigma \lambda + \frac{\sigma \lambda^2}{2}
\end{equation}
in this case.



\section{Non convexity close to the transition \label{sec:app:non-convexity}}

\begin{figure}
\begin{center}
\includegraphics[width=0.45\textwidth]{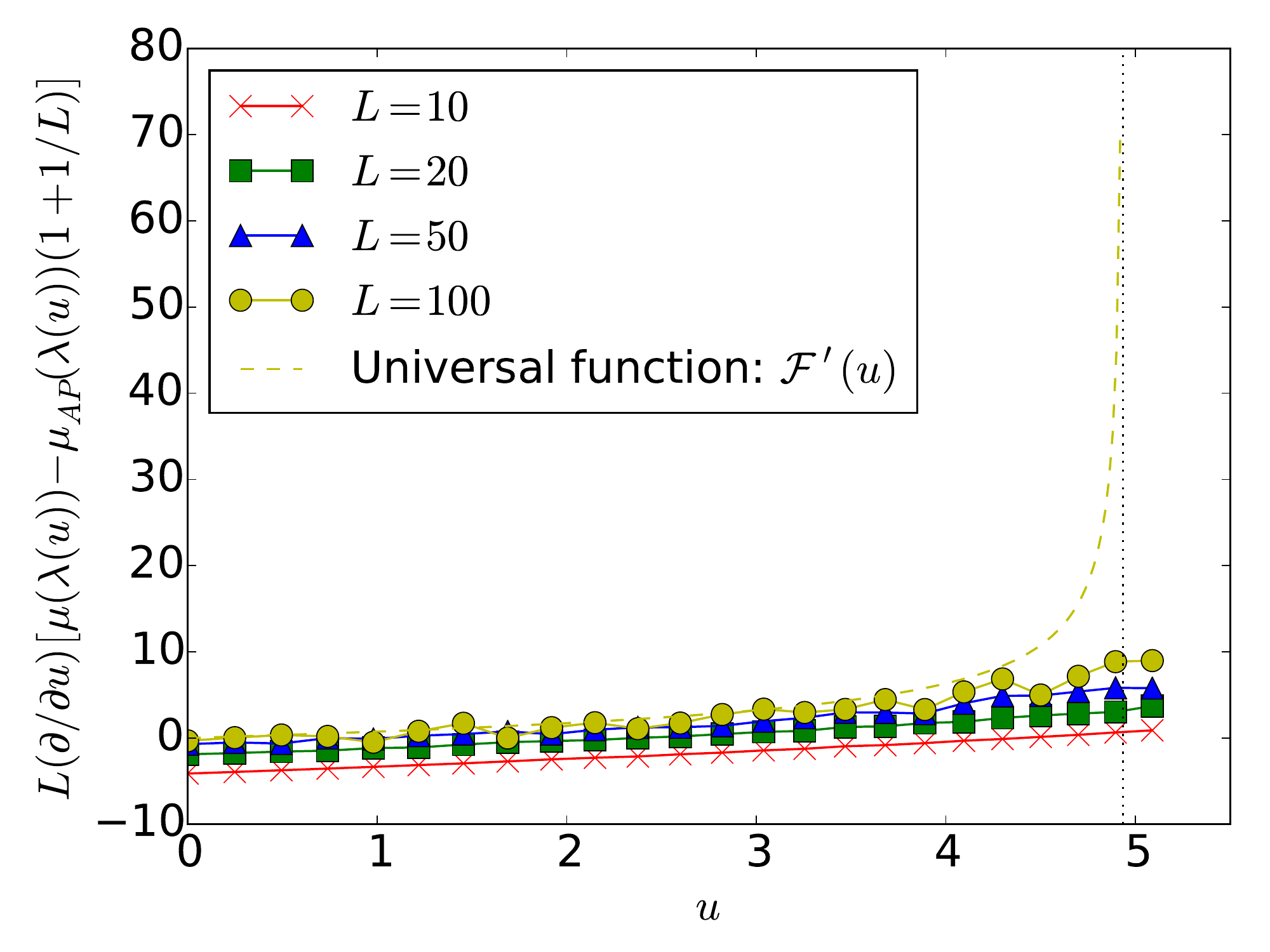} 
\caption{\label{fig:Universal} 
$L(\partial/\partial u)[\mu(\lambda)-\mu_{\rm AP}(\lambda(u)) (1+1/L)]$ for the WASEP close to $u^\star$, where $\mu(\lambda)$ is estimated using the cloning algorithm 
\cite{Giardina2006xxx}.
Different markers represent different values of $L$.
Away from the transition point $u^\star$, the $\frac{1}{L}$ correction $\mathcal F^{\prime}$ correctly captures the first derivative of the cumulant generating function, as predicted 
~\cite{Appert-Rolland2008}.
As  $u \rightarrow u^\star$, the derivative of ${\mathcal F}(u)$ diverges. 
It is clear that ${\mathcal F}(u)$ does not capture the finite size corrections near the transition. 
}
\end{center}
\end{figure}

For the WASEP, we show numerical examples of $L(\partial/\partial u)[\mu(\lambda)-\mu_{\rm AP}(\lambda(u)) (1+1/L)]$ in Fig.~\ref{fig:Universal}, which converges to $\mathcal F'(u)$ away from the transition point $u^*=\pi^2 / 2$ 
~\cite{Appert-Rolland2008}.
Close to the transition point, 
although $\mathcal F'(u)$ diverges at $u^{\star}$, there is no sign of the precursor of the corresponding divergence in numerics up to $L=100$ in the figure. This corroborates that $\mathcal F(u)$ does not describe the CGF (LDF) close to the transition $u^{\star}$.

Aside from the divergence of the higher order corrections, let us argue that $\mathcal F(u)$ cannot describe  the LDF close to the transition point. It is safe to assume the existence and the differentiability of $\mu$ in the domain around $u^{\star}$ for finite,but large, system size $L$. Indeed, this is what we observe numerically as seen in Fig.~\ref{fig:Universal} for example.
From the Gartner-Ellis theorem (see e.g. 
\cite{Touchette2009} 
and references therein), 
we find that the corresponding LDF exists and 
the LDF needs to be convex function. However, to order $1/L$,  the LDF  becomes non-convex for $u\rightarrow u^\star$ for any finite $L$. This can be seen from the negative divergence of the second derivative of $\mathcal{F}$ close to the transition point. Therefore,  the first order perturbation eq.\eqref{eq:Phi1} 
in the main text cannot describe the correct behavior of the LDF close to the transition.

This violation of the first order perturbation implies that higher order corrections also need to diverge as a compensation.  Resumming all the corrections restores the convexity below the transition.

\section{Finite size corrections for the SSEP on a ring
\label{sec:app:periodic corr}}
A powerful method to study finite size corrections in integrable spin chain models is based on the so-called Baxter equation 
\cite{Gromov:2005gp}
\begin{equation} \label{eq:Baxter}
\mathcal{T}(z)  = e^{\frac{\lambda}{2}}\left(z+\frac{i}{2}\right)^{L}\frac{ \mathcal{Q}(z-i)}{\mathcal{Q}(z)}+ e^{-\frac{\lambda}{2}}\left(z-\frac{i}{2}\right)^{L} \frac{\mathcal{Q}(z+i)}{\mathcal{Q}(z)}\,,
\end{equation}
where $\mathcal{Q}(z)=\prod_{j=1}^{N}(z-z_j)$ is the Baxter polynomial. $\mathcal{T}(z)$ is the transfer matrix which is a polynomial of degree $L$  whose explicit form will not be important here but it can be fixed by self-consistency of the equation. 
The Bethe equations 
\eqref{eq:XXX} 
follow from \eqref{eq:Baxter} by requiring that the residues in the right hand side at the location of the zeros of $\mathcal{Q}(z)$ vanish.

We will now consider the thermodynamic limit of the Baxter equation as $N,L \rightarrow \infty$ with $z_j \sim \mathcal{O}(L)$ following the method introduced in 
\cite{Gromov:2005gp}.
In this limit, it is  useful to introduce the rescaled rapidity $w$ as $w\equiv z/L$ and use the notations $\varphi(w) \equiv \frac{1}{L}\sum_{j=1}^{N} \log(w-w_k)$, the potential $V(w)\equiv\log(w)$ and the rescaled transfer matrix $t(w) \equiv \mathcal{T}(L w)/(2 (L w)^{L})$, where all these functions are now  of $\mathcal{O}(1)$. In this notation, the Baxter equation looks like
\begin{equation} \label{eq:thermobaxter}
t(w) =\frac{1}{2}\left(\exp \left( L\, \Lambda^{+}(w)+\frac{\lambda}{2} \right)+\exp \left( L\,\Lambda^{-}(w)-\frac{\lambda}{2} \right) \right)
\end{equation}
with
\begin{equation}
\Lambda^{\pm}(w) \equiv \varphi\left(w \mp\frac{i}{L}\right) - \varphi(w)+V\left(w\pm\frac{i}{2L}\right)- V(w)  \,.
\end{equation}
In addition, we introduce the \textit{quasi-momentum} $p(w)$ which will play a major role in the subsequent analysis as
\begin{equation} \label{eq:quasimomentum}
p(w) \equiv \varphi'(w) - \frac{V'(w)}{2}+\frac{i\lambda}{2}\,.
\end{equation}
By definition, the expansion of $p'(w)$ around $w=0$ is given by
\begin{equation}\label{energy}
p'(w) -\frac{1}{2w^2} \stackrel{w\rightarrow 0}{\simeq}  - \frac{1}{L}\sum_{j=1}^{N}\frac{1}{w_j} =  G \,.
\end{equation}
The idea is now to expand the Baxter equation in $1/L$ and solve it for $p(w)$, order by order in this parameter. The first few orders look as follows,
\begin{equation}
\begin{aligned}
t(w)&=\cos (p(w)) \left(1-\frac{4 p'(w)+3 V''(w)}{8 L}\right)+\mathcal{O}(1/L^2)\,.
\end{aligned}
\end{equation}
We aim at finding the first corrections to $p(w)$ and hence, we will expand both $p(w)$ and $t(w)$ in powers of $1/L$ as 
\begin{equation}
\begin{aligned}
&p(w) = p_0(w)+\frac{1}{L} p_1(w) +\mathcal{O}(1/L^2) \\
& t(w) = t_0(w)+\frac{1}{L} t_1(w)+\mathcal{O}(1/L^2)\,.
\end{aligned}
\end{equation}
The polynomiality condition on the transfer matrix $\mathcal{T}(z)$ lead us to assume that the coefficients $t_i$ contain only one singularity at $w=0$ (by construction) but apart from that they are analytic everywhere in $w$.

At each order, we will have a Riemann-Hilbert problem defining the quasi-momentum whose asymptotic conditions are imposed directly from its definition (\ref{eq:quasimomentum}) and the analytic properties follow from the properties of the functions involved in (\ref{eq:thermobaxter}).

\subsection{Leading order correction}
Let us gain some intuition of what happens to the Bethe roots when we consider the thermodynamic limit at the level of the Bethe equations (\ref{BAE}). Taking the logarithm of these equations we get that
\begin{equation}\label{BAE}
L \log\left(\frac{z_j +\frac{i}{2}}{z_j -\frac{i}{2}} \right) = \sum_{j=1}^{N} \log\left(\frac{z_i-z_j +i}{z_i-z_j - i} \right) +2 \pi i \left(n_j +  \frac{i\lambda}{2\pi}\right)\,,
\end{equation} 
where $n_j$ is an integer called mode number that parametrizes the branches of the logarithms and the twist $\lambda$ corresponds to a shift of such mode number. As we go to the thermodynamic limit it is well known that the Bethe roots condense into cuts in the complex plane whose position depends on the choice of the mode number $n_j$. The presence of the twist will merely change the shape of the cuts as well as their positions. In order to specify for the ground state, we use the observation made in 
\cite{Appert-Rolland2008}
that the distribution of Bethe roots is along a single contour (often called a one-cut solution) where $n_j = n$ for all $j$ and as we will see we further need to set $n$ to zero.

In terms of the quasi-momentum, the thermodynamic limit of the equation (\ref{BAE}) reads 
\cite{Gromov:2005gp}
\begin{equation}
p_0(z+i 0)+p_0(z-i0)= 2 \pi n_j\,\,\, \text{for}\,\,\, z\in C_j\,,
\end{equation}
where $C_j$ denotes the cut in the complex plane formed by the Bethe roots. Considering the derivative of the quasi-momentum $p'(z)$, we see that it satisfies
\begin{equation}
p_0'(z+i0) = -p_0'(z-i0)\,,
\end{equation}
which means that $\pm p_0'(z)$ defines a two-sheeted Riemann surface with square-root type branch points.

To see this from the point of view of the equation (\ref{eq:thermobaxter}), we have that at leading order
\begin{equation}
t_0(z) = \cos (p_{0}(z))\,,
\end{equation}
which implies that the derivative of the quasi-momentum at this order will be given by
\begin{equation}
p'_{0}(z)=\frac{t'_0(z)}{\sqrt{1-t_0^{2}(z)}}\,.
\end{equation}
Given the analyticity of $t_0$ we see explicitly the square-root branch points emerging from the denominator of the expression above.
Using the definition of the quasi-momentum (\ref{eq:quasimomentum}), we can supplement the previous equation with the  asymptotic condition 
\begin{equation}
p_0'(z) \sim \frac{1}{2 z^2}+\mathcal{O}(1)\,\,\text{as}\,\, z\rightarrow 0\,,
\end{equation}
and
\begin{equation}
p_0'(z) \sim \frac{1}{z^2}\left(\frac{1}{2}-\rho\right)\,\,\text{as}\,\, z\rightarrow \infty\,, 
\end{equation}
with $\rho = N/L$ being the density of particles. 
This fixes $p_0'(z)$ to be
\begin{equation}
p_0'(z)=\frac{2 \theta_1 \theta_2-z (\theta_1+\theta_2)}{4 z^2 \sqrt{\theta_1 \theta_2} \sqrt{(z-\theta_1) (z-\theta_2)}}\,,
\end{equation}
where $\theta_1$ and $\theta_2$ are the branch points where the distribution of the Bethe roots ends. We can relate them back to the rapidity variables $\xi$ of the original Bethe equations 
and replace the location of the branch points found in 
\cite{Appert-Rolland2008}
namely $\xi=e^{-is(\pm \theta+2i \rho)}$ with $\theta=2 \sqrt{\rho(1-\rho)}$. Using (\ref{energy}), we then obtain the leading term for the energy to be
\begin{equation}
\mu_0=\frac{\theta ^2 \lambda ^2}{4}\,,
\end{equation}
which matches the result found in 
\cite{Appert-Rolland2008}.

\subsection{Subleading corrections}
\begin{figure}
\begin{center}
\includegraphics[width=0.4\textwidth]{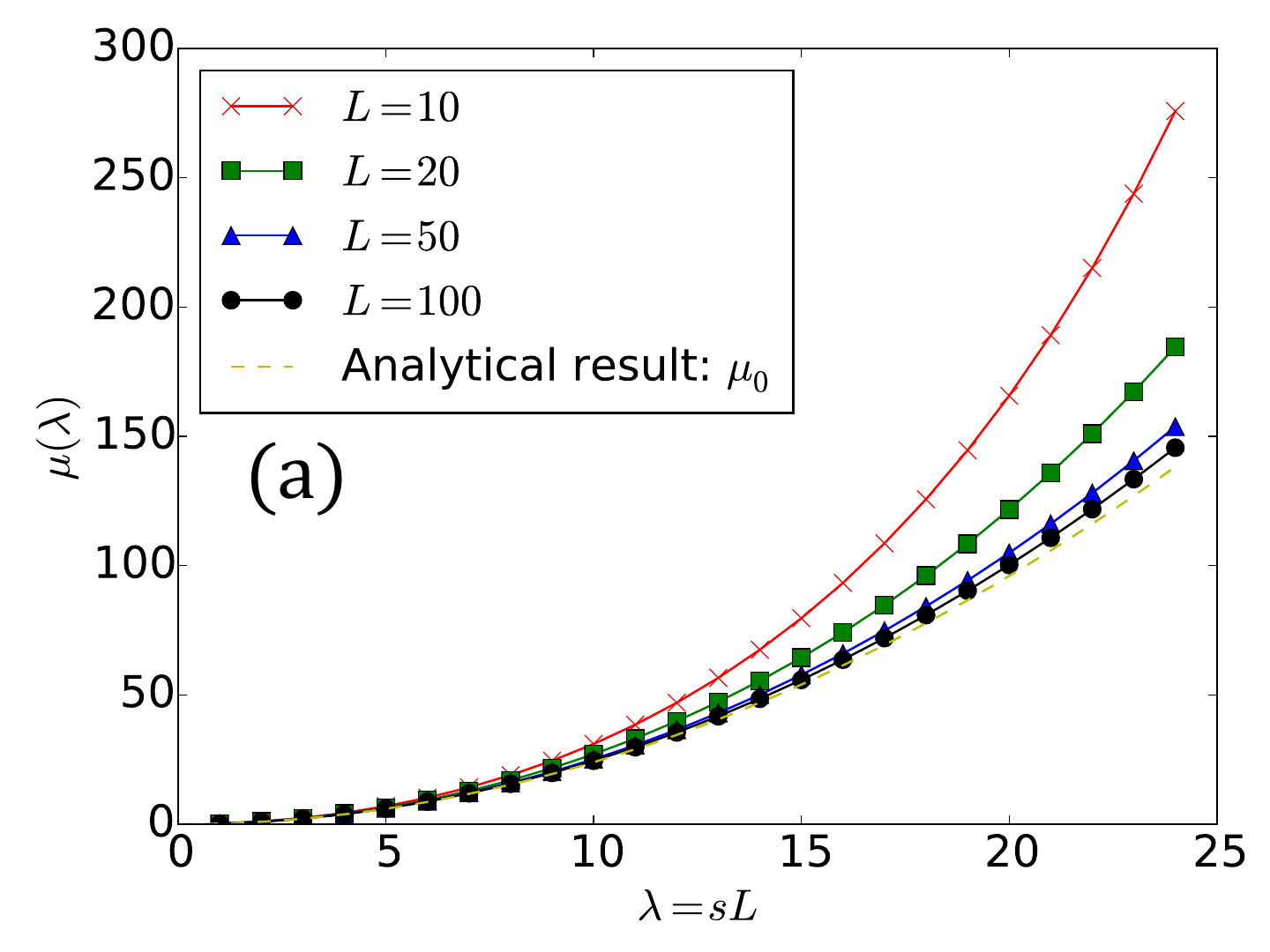} 
\includegraphics[width=0.4\textwidth]{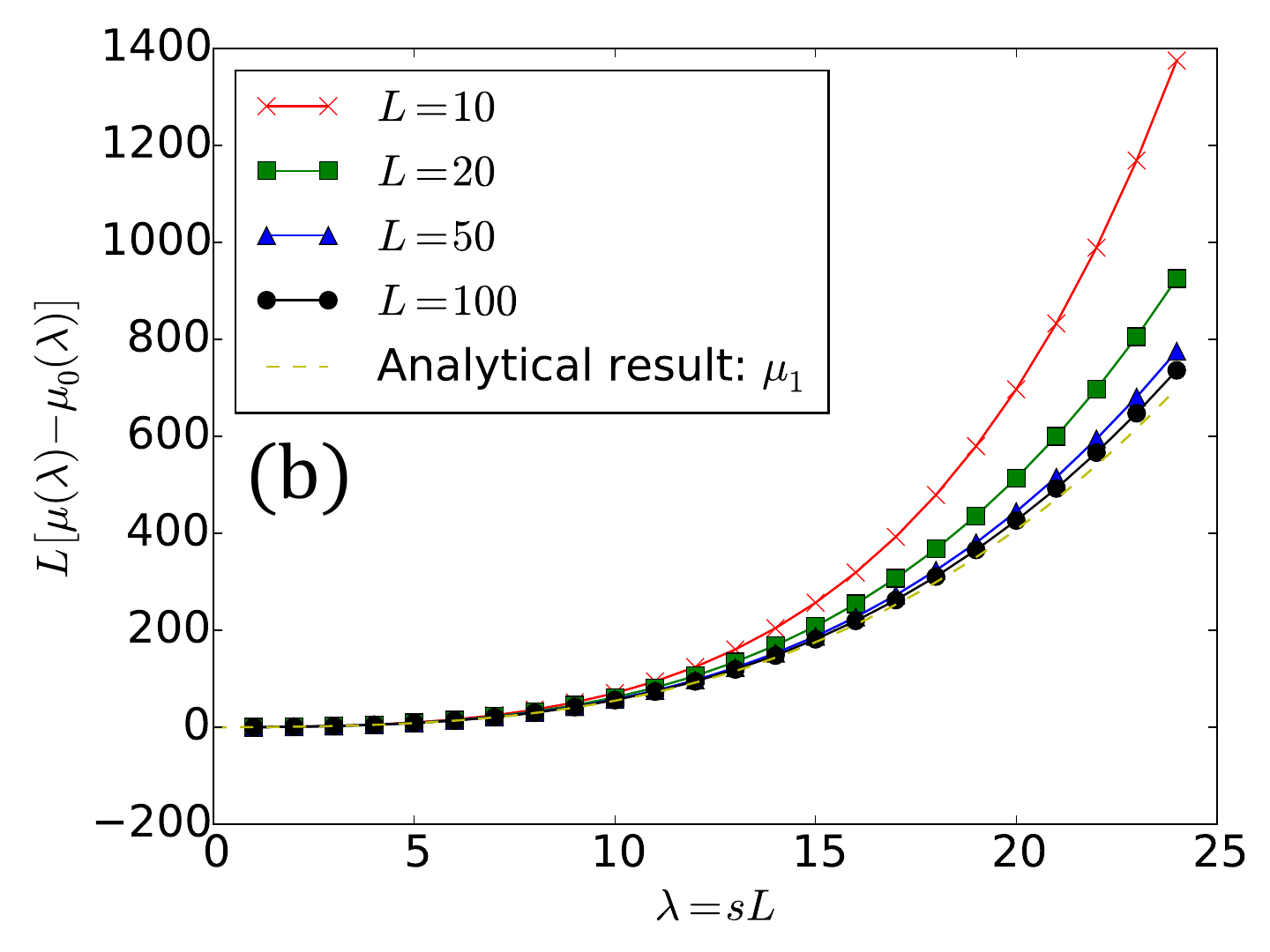} 
\includegraphics[width=0.4\textwidth]{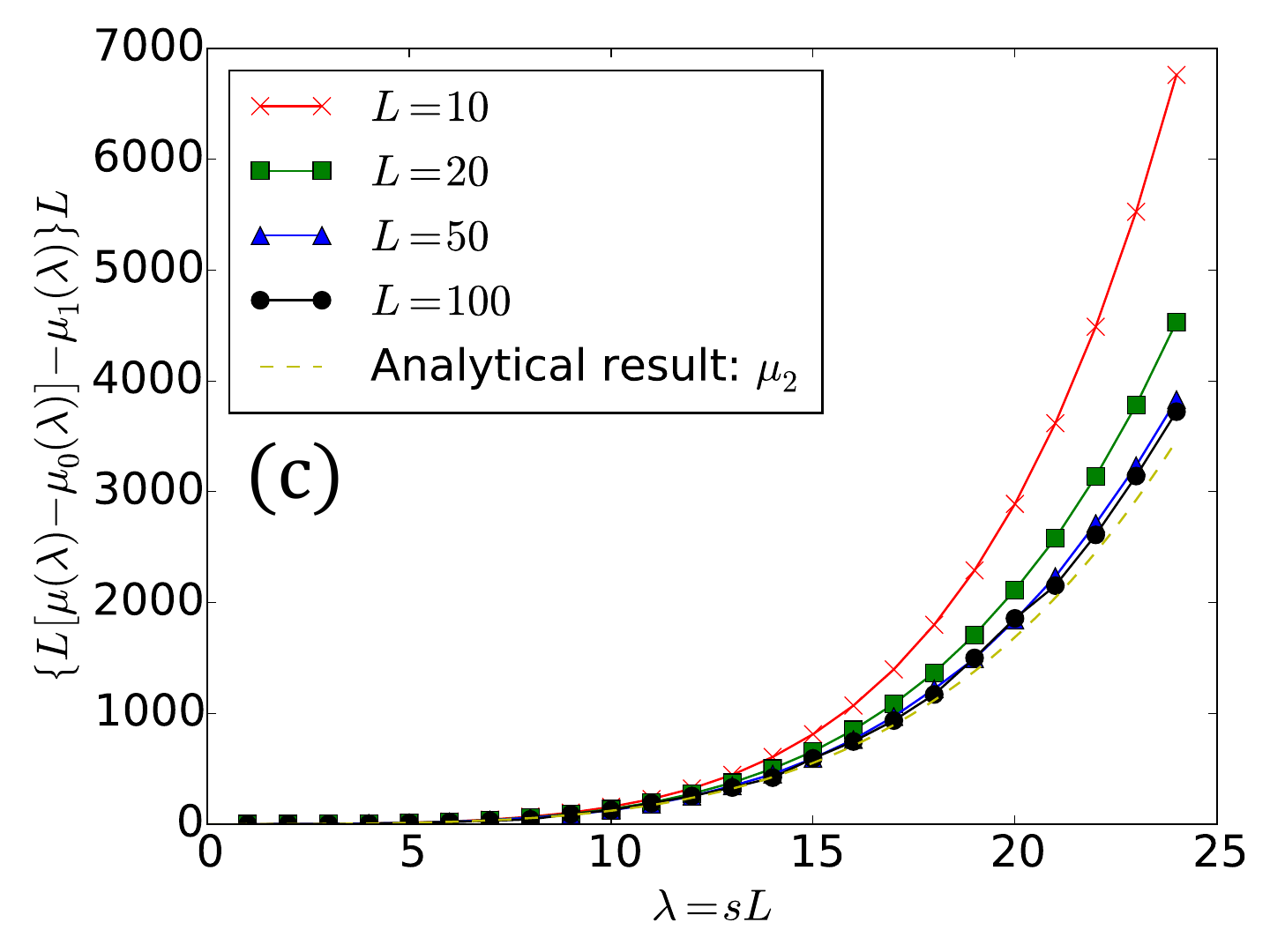} 
\caption{\label{fig:SSEP_ThirdOrder} Analytical results of $\mu_{0}$, $\mu_{1}$ \eqref{SM:mu0_1analy} and $\mu_{2}$ \eqref{eq:mu2_exact} for $\rho = 0.4$ are shown as yellow dashed lines in the panel (a), (b) and (c), respectively. We also plot the corresponding numerical results obtained from the cloning algorithm 
\cite{Giardina2006xxx} 
for several values of $L$. We can see the convergence of numerical results to the corresponding analytical predictions as $L\rightarrow \infty$. }
\end{center}
\end{figure}

We could proceed in expanding one order further the equation~(\ref{eq:thermobaxter}) and solve the corresponding Riemann-Hilbert problem. Instead, we will use the results already available in literature derived by this method in 
\cite{Gromov:2005gp} 
to recover the  already known next-to-leading order correction and make a new prediction for the next-to-next-to-leading order. The result of 
\cite{Gromov:2005gp} 
was derived for a closely related spin chain model, namely the (untwisted) $sl(2)$ spin chain, whose Bethe equations are related to those in 
\eqref{eq:BAE}
simply by replacing $L\rightarrow-L$. Additionally we have seen that the twist shifts the mode numbers. The one-cut solution of the leading and next-to-leading correction to the energy in 
\cite{Gromov:2005gp,Beisert:2005mq}
for the $sl(2)$ spin chain reads
\begin{equation}
\begin{aligned}
\mu_0=&-4 \pi ^2 n^2 \rho  (\rho +1)\\
\mu_1=&-4 \pi ^2\Biggl(\sum_{k=1}^{\infty}\biggl[k \sqrt{k^2+4 n^2 \rho  (\rho +1)}-k^2\\
&-2 n^2 \rho  (\rho +1)\biggr]-n^2 \rho  (\rho +1)\Biggr)
\end{aligned}
\label{SM:mu0_1analy}
\end{equation}
Upon the transforming $L\rightarrow -L$ (and thus $\rho \rightarrow -\rho$) as well as $n\rightarrow n+ \frac{i\lambda}{2\pi}$  we obtain the complete result of 
\cite{Appert-Rolland2008}
when we finally set $n=0$ (which corresponds to selecting the ground state).
This encourages us to proceed to the next-to-next-to-leading order and apply the same heuristic rule to the result written in the appendix B of 
\cite{Gromov:2005gp} 
in order to obtain the model under study. Let us first define the following auxiliary sums,
\begin{equation}
\begin{aligned}
& \mathcal{S}_1 = \sum_{l=1}^{\infty} \frac{\sqrt{\theta ^2 \lambda ^2+4 \pi ^2 l^2}}{2 \pi  l}-1 \\
&\mathcal{S}_2 = \sum_{l=1}^{\infty}\left(-\frac{\theta ^2 \lambda ^2}{8 \pi ^2}+\frac{l}{2\pi} \sqrt{\theta ^2 \lambda ^2+4 \pi ^2 l^2}-l^2\right)\\
&\mathcal{S}_3(k)= \sum_{l=1}^{\infty}\frac{k \sqrt{\theta ^2 \lambda ^2+4 \pi ^2 k^2}-l \sqrt{\theta ^2 \lambda ^2+4 \pi ^2 l^2}}{2 \pi  k^2-2 \pi  l^2}-1\,,
\end{aligned}
\end{equation}
we have that the correction reads
\begin{equation}
\begin{aligned}
\label{eq:mu2_exact}
\mu_2=&\frac{\lambda ^2}{96}  \biggl(\theta ^2 \left(\left(19 \theta ^2-14\right) \lambda ^2+24\right)\\
&+96 \left(4 \theta ^2-3\right) (\mathcal{S}_1^2-\mathcal{S}_1) \biggr)+8 \pi ^2 \mathcal{S}_2\\
&-\sum_{k=1}^{\infty}\biggl(2 \left(3 \theta ^2-2\right) \lambda ^2+\frac{\theta ^2 \left(4 \theta ^2-3\right) \lambda ^4}{4 \pi ^2 k^2}\\
&+8 \pi ^2 \left(k^2+2 \mathcal{S}_2\right)\\
&+\frac{(2 \mathcal{S}_3(k)-1) \left(\theta ^2 \left(4 \theta ^2-3\right) \lambda ^4+32 \pi ^4 k^4\right)}{2 \pi  k \sqrt{\theta ^2 \lambda ^2+4 \pi ^2 k^2}} \\
&+\frac{(2 \mathcal{S}_3(k)-1) \left(4 \pi ^2 \left(7 \theta ^2-4\right) \lambda ^2 k^2\right)}{2 \pi  k \sqrt{\theta ^2 \lambda ^2+4 \pi ^2 k^2}} \biggr)\,. 
\end{aligned}
\end{equation}
The sums above mimic the sums over Fourier modes of the wave fluctuations in the MFT. We verified this result numerically via the cloning algorithm, see Fig.~\ref{fig:SSEP_ThirdOrder}.

\section{Finite size corrections using the MFT for boundary driven system 
\label{sec:app:driven corr }}

Let us bring in full details, the finite size corrections to the LDF $\Phi(J)$ to order $1/L^2$.

First, the $S_3,S_4$ terms are given by 
\begin{eqnarray}
S_{3} = \sum_{\vec{k}, \vec{\omega}} \chi_0 h_0 V_3  & \quad \quad& 
S_4 = \sum_{\vec{k}, \vec{\omega}} \sum_{l=1,2,3} \chi_l g_l  V_4 
\end{eqnarray}
where  $V_m =  \prod_{j=1,...,m} a_{k_j,\omega_j} \delta \left( \sum k_i\right) \delta \left(\sum \omega_i\right)
$ and 
\begin{eqnarray}
\chi_0 = -\frac{J \sigma''}{2\sigma^2}  & \quad \quad &
\chi_1 = \frac{DD''}{2\sigma}
 \\ \nonumber
 \chi_2 = \frac{J^2 (\sigma'') ^2}{8\sigma^3} &+& \frac{\sigma^{(4)}}{48}(E^2-J^2/\sigma^2)  
 \\ \nonumber 
 \chi_3 = -\frac{\sigma''}{4\sigma^2}
&\quad \quad &
 h_0 = -i \omega_3 k_1 k_2 
 \\ \nonumber 
 g_1 = -k_1 k_2 k_3 ^2 k_4 ^2 
&\quad \quad&
 g_2 = k_1 k_2 k_3  k_4  
 \\ \nonumber 
 g_3 &=& k_1 k_2 \omega_3 \omega_4 
\end{eqnarray}

Performing the perturbative expansion to order $1/L^2$, we find
\begin{eqnarray}
\label{eq:Phi2 full}
\Phi_2 &=& 
\frac{15\sigma^{2}}{4D^{2}}
\left(\chi_{1}u+\frac{3}{2}\chi_{2}\right)\left(1-\sqrt{2u}\cot\sqrt{2u}
\right)
\\ \nonumber & &
\frac{15\sigma^{2}}{4D^{2}}
\left(\chi_{1}\mathcal{G}_{1}\mathfrak{\mathcal{G}}_{3}-2\chi_{1}\mathcal{G}_{2}^{2}+3\chi_{2}\mathcal{G}_{1}^{2} \right) 
\\ \nonumber & &
+ \frac{2\sigma^{3}}{D^{2}}\chi_{0}^{2}\mathcal{G}_{2}
+\frac{2\chi_{0}^{2}}{\pi^{2}}\left( F_{a}\left(u\right) + F_{b}\left(u\right)\right)
\\ \nonumber & &
-\frac{15\sigma^{2}}{2\pi^{2}}\chi_{3}\left(\mathcal{H}_{1}+\mathcal{H}_{2}\right)
\\ \nonumber & &
 + \left( A u + B u^2 \right) (1 + \frac{C}{\sqrt(\delta u )} ),  
\end{eqnarray}

where $A,B,C$ are non-universal terms that depend on the cutoffs and with the functions  
\begin{eqnarray}
F_{a}\left(u\right)&=&\sum_{k_{1},k_{2}} \int d\omega_{1}d\omega_{2}\,\frac{k_{1}k_{2}^{3} \omega_{1}\left(\omega_{1}+\omega_{2}\right)^{2}}{f\left(k_{1},\omega_{1}\right)f\left(k_{2},\omega_{2}\right)f\left(k_{1}+k_{2},\omega_{1}+\omega_{2}\right)}
  \nonumber \\  \nonumber 
F_{b}\left(u\right)&=&\sum_{k_{1},k_{2}}\int d\omega_{1}d\omega_{2}\,\frac{k_{1}^{2}k_{2}^{2}\left(\omega_{1}+\omega_{2}\right)^{2}}{f\left(k_{1},\omega_{1}\right)f\left(k_{2},\omega_{2}\right)f\left(k_{1}+k_{2},\omega_{1}+\omega_{2}\right)}
 \\ \nonumber 
 \mathcal{G}_{1}&=&\sum_{k}\frac{k}{\sqrt{k^{2}-2u}}-1
  \\ 
\mathcal{G}_{2}&=&\sum_{k}\frac{k^{2}}{\sqrt{k^{2}-2u}}-k-\frac{u}{k}
 \\ \nonumber 
\mathcal{G}_{3}&=&\sum_{k}\frac{k^{3}}{\sqrt{k^{2}-2u}}-k^{2}-u
 \\ \nonumber 
\mathcal{H}_{1}&=&8\sum_{k}k^{2}\log Dk\log\left(k^{2}-2u\right)
 \\ \nonumber & & 
 \quad \quad -k^{2}\log Dk\log k^{2}+2u\log Dk 
 \\ \nonumber 
\mathcal{H}_{2}&=&\sum_{k}k^{2}\log^{2}\left(k^{2}-2u\right)-4k^{2}\log^{2}k+8u\log k .
\end{eqnarray}

One can show that as $\delta u \rightarrow 0$, only the first two lines of \eqref{eq:Phi2 full} contribute to the leading $1/\delta u$ divergence.


\end{document}